\documentclass[useAMS,usenatbib]{mn2e}
\usepackage{amsfonts}
\usepackage{bbm}
\usepackage{graphicx}

\newcommand{\kms}{\,{\rm km\,s^{-1}}}

\title[A New model for the Galactic Bar]
{A New Model for the Milky Way Bar}
\author[Y. Wang et al.]
{Yougang Wang$^{1,2}$\thanks{email: wangyg@bao.ac.cn},
Hongsheng Zhao$^{2,3,4}$, Shude Mao$^{2,5}$, R. M. Rich$^{6}$ \\
$^1$Key Laboratory of Optical Astronomy, National Astronomical
Observatories, Chinese
Academy of Sciences, Beijing 100012, China\\
$^2$National Astronomical Observatories, Chinese
Academy of Sciences, Beijing 100012, China\\
$^3$SUPA, University of St Andrews, KY16 9SS, UK\\
$^4$Vrije Universiteit, De Boelelaan 1081, 1081 HV Amsterdam, The
Netherlands\\
$^5$Jodrell Bank Centre for Astrophysics, University of Manchester, Manchester M13 9PL, UK  \\
$^6$Department of Physics and Astronomy, University of California,
Los Angeles, CA 90095-1562, USA}
\begin{document}

\date{Accepted . Received .}

\pagerange{\pageref{firstpage}--\pageref{lastpage}} \pubyear{2012}

\maketitle

\label{firstpage}

\begin{abstract}
We use Schwarzschild's orbit-superposition technique to construct
self-consistent models of the Galactic bar. Using $\chi^2$
minimisation, we find that the best-fit Galactic bar model has a
pattern speed $\Omega_{\rm p}=60\ \rm{km\ s^{-1}\ kpc^{-1}}$, disk
mass $\rm{M_{\rm d}=1.0\times10^{11}M_{\odot}}$ and bar angle
$\theta_{\rm bar}=20^{\circ}$ for an adopted bar mass $\rm{M_{\rm
bar}=2\times10^{10}M_{\odot}}$. The model can reproduce not only the
three-dimensional and projected density distributions but also
velocity and velocity dispersion data from the BRAVA survey. We also
predict the proper motions in the range
$l=[-12^{\circ},12^{\circ}]$, $b=[-10^{\circ},10^{\circ}]$, which
appear to be higher than observations in the longitudinal direction.
The model is stable within a timescale of 0.5 Gyr, but appears to
deviate from steady-state on longer timescales. Our model can be
further tested by future observations such as those from GAIA.
\end{abstract}

\section{Introduction}
It is well known that most spiral galaxies host a bar structure in
their central region \citep[e.g.][]{2011arXiv1110.1933L}. Therefore,
one of the most important issues in galaxy formation and evolution
is to understand the structure and dynamical properties of barred
galaxies. Our own Milky Way is the nearest barred galaxy. Compared
with other distant galaxies, the Galaxy has extensive observed
photometric and kinematic data which enable us to study the bar
structure in detail. A thorough understanding of the structure and
dynamics of the Galactic bar may help us understand the formation of
other spiral galaxies, and test the validity of the popular Cold
Dark Matter structure formation model
\citep[e.g.][]{2010ApJ...720L..72S}.

Observationally, the Galactic bar model can be constrained by
surface brightness maps
\citep{1995ApJ...445..716D,2005MNRAS.358.1309B,2011A&A...534L..14G},
microlensing optical depth maps
\citep{1991ApJ...379..631B,1995ApJ...440L..13Z} and star counts
\citep{1994ApJ...429L..73S,2002MNRAS.337..895M,2007MNRAS.378.1064R}.
Different observational techniques, wavelengths and fields probe
different aspects of the bar. Schwarzschild's orbit-superposition
technique \citep{1979ApJ...232..236S} provides us the possibility to
construct a model that can fit all observations. Using the
orbit-superposition technique, Zhao (1996, hereafter ZH96)
constructed a self-consistent Galactic bar model, which can fit the
surface brightness, velocity and velocity dispersion in the Baade's
window (BW). However, recently it was found that the predicted
rotation curve in ZH96 is inconsistent with the results from the
Bulge Radial Velocity Assay
\citep[BRAVA,][]{2007ApJ...658L..29R,2008ApJ...688.1060H,2012AJ....143...57K}.
Possible explanations are (1) the data resolution in ZH96 is not
sufficiently high, and/or (2) the initial conditions of orbits
include too many loop orbits. Also using the orbit-superposition
technique, \cite{2000MNRAS.314..433H} constructed a dynamical model
of the inner Galaxy, which fits most of the available data in one
very intensively observed bulge field, Baade's Window
$(l,b)=(0,-4)$; the only disadvantage is that the proper motion
dispersion in their model is higher than the measured values by
\cite{1992AJ....103..297S}. Another useful method to generate
self-consistent dynamical models is the made-to-measure algorithm
\citep{1996MNRAS.282..223S,2007MNRAS.376...71D,2009MNRAS.395.1079D,2010MNRAS.405..301L},
which was implemented for the Milk Way by
\cite{2004ApJ...601L.155B}. However, only the density map of Milk
Way was used to construct the dynamical model, no kinematic
constraints were applied and their effective field is small.
Furthermore, the effective particle number in their study turns out
to be small (only a few thousand, \citealt{2007MNRAS.378.1064R}).

Recently, extensive observations of the central region of the Milky
Way by the Hubble Space Telescope \citep{2006MNRAS.370..435K} and
ground-based telescopes (BRAVA:
\citealt{2007ApJ...658L..29R,2008ApJ...688.1060H,2012AJ....143...57K};
OGLE: \citealt{2000AcA....50....1U,2004MNRAS.348.1439S}) provide
many other large samples of kinematic data. In this paper, our aim
is to construct a self-consistent and stable bar model which can fit
all the presently-available observed data in the central region of
the Milky Way by using the Schwarzschild method. As we will see
later, we are able to produce a self-consistent model, but there are
issues with long term stability.

The paper is organized as follows. In section 2, we describe the
density and potential model for this study. Section 3 presents the
details of our implementation of the Schwarzschild method. In
section 4, we show the main results. Section 5 shows the stability
of the bar model. Conclusion and discussion are given in section 6.
Throughout this paper, we adopt the distance to the Galactic center
as $R_{\rm 0}=8\ {\rm kpc}$.

\section{Density and Potential of the central region of the Galaxy}

\begin{figure}
\hspace{-1.5cm}
\includegraphics[angle=0, width=100mm]{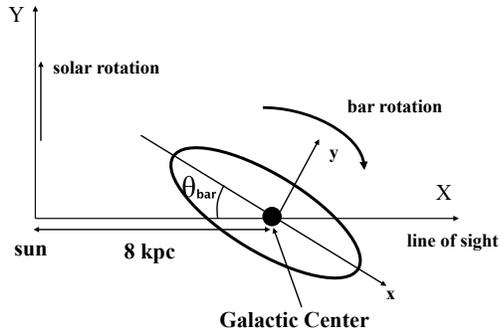}
\vspace{-1.5cm}
\caption{The coordinate system used throughout the paper. The major
axis ($x$-axis) of the bar is misaligned with the line of sight
($X$-axis) with an angle $\theta_{\rm bar}$. The $z$-axis comes out
of the page. The bar rotation and the rotation of the Sun with
respect to the Milky Way centre are indicated. Notice the rotation
of the bar is in the negative $z$-axis and the near-side of the bar has positive longitude ($l$). } \label{fig:coordinate}
\end{figure}

For clarity, in Figure~\ref{fig:coordinate}, we first establish the
coordinate system we use throughout this paper. In particular, the
major axis of the bar is along the $x$-axis, and is at an angle
$\theta_{\rm bar}$ with respective to the line of sight (the
$X$-axis). Notice that the bar angular momentum is along the
negative $z$-axis, but we still write the pattern speed as positive
for abbreviation.

\subsection{Density and potential of the bar and bulge}

 Lots of observation data have been used to constrain
dynamical models of the Galaxy. Using the \emph{COBE} DIRBE
multiwavelength observations, \cite{1995ApJ...445..716D} constructed
many analytic bar models. \cite{1997ApJ...477..163S},
\cite{2005MNRAS.358.1309B} and \cite{2007MNRAS.378.1064R} used the
star counts of the red clump giant stars to constrain the triaxial
Galactic bar models. Both \cite{1997ApJ...477..163S} and
\cite{2007MNRAS.378.1064R} found the analytic bar model given in
\cite{1995ApJ...445..716D} fit the data well. In this paper, we also
adopt the \cite{1995ApJ...445..716D} bar model as in ZH96, which has
the following form
\begin{equation}
\rho(x,y,z)=\rho_0\bigg[{\exp}\bigg(-\frac{r_1^2}{2}\bigg)+r_2^{-1.85}{\exp}(-r_2)\bigg],
\end{equation}
where the first term represents a bar with a Gaussian radial profile
and the second term a spheroidal nucleus with a steep inner power
law and an exponential outer profile. In this paper, we refer
both the bar and nuclear components simply as the ``bar''. The
central density $\rho_0$ is determined by normalising the total mass
of the bar, $M_{\rm bar}=2.0\times 10^{10} M_\odot$, which is fixed
throughout the paper. The radial functions $r_1$ and $r_2$ are
defined as
\begin{equation}
r_1=\bigg\{\bigg[\bigg(\frac{x}{x_0}\bigg)^2+\bigg(\frac{y}{y_0}\bigg)^2
\bigg]^2+\bigg(\frac{z}{z_0}\bigg)^4 \bigg\}^{1/4}
\end{equation}
and
\begin{equation}
r_2=\bigg[\frac{q^2(x^2+y^2)+z^2}{z_0^2}\bigg]^{1/2},
\end{equation}
where the principal axes of the bar are $x_0=1.49\ \rm{kpc}$,
$y_0=0.58\ \rm{kpc}$, $z_0=0.40\ \rm{kpc}$ and the bulge axis ratio
is $q=0.6$.

\subsection{Disc potential and density}
We do not include explicitly any dark halo in our potential.
\cite{2002ApJ...573..597K} shows that the dark halo has to be very
low in mass in the central part in order to allow many microlensing
events by baryonic material. Instead of the usual halo plus
exponential disk, we use a Miyamoto-Nagai potential to represent the
disk plus halo. The disk potential is given by
\begin{equation}
\Phi_d(x,y,z)=-\frac{GM_d}{r_3},
\end{equation}
where
\begin{equation}
r_3=\bigg\{x^2+y^2+\bigg[a_{\rm{MN}}+(z^2+b_{\rm{MN}}^2)^{1/2}\bigg]^2
\bigg\}^{1/2},
\end{equation}
$a_{\rm{MN}}=6.5$ kpc, $b_{\rm{MN}}=0.26$ kpc and $M_{\rm{d}}$ is
the total disc mass. In ZH06, $M_{\rm{d}}=8M_{\rm{bar}}$ and
$M_{\rm{bar}}=2\times10^{10}M_{\odot}$ is the total mass of the bar.
In this paper, we will consider different values of the disk mass but keep the bar mass fixed.

The density distribution of the MN disk is given by
\begin{eqnarray}
&&\rho_{\rm d}(x,y,z)=\bigg(\frac{b_{\rm{MN}}^2M_d}{4\pi}\bigg)\nonumber\\
&&\times\frac{a_{\rm{MN}}R^2+(a_{\rm{MN}}+3\sqrt{z^2+b_{\rm{MN}}^2})(a_{\rm{MN}}+\sqrt{z^2+b_{\rm{MN}}^2})^2}{[R^2+(a_{\rm{MN}}+\sqrt{z^2+b_{\rm{MN}}^2})^2]^{5/2}(z^2+b_{\rm{MN}}^2)^{3/2}} \nonumber
\end{eqnarray}
where $R^2=x^2+y^2$.

\subsection{Density, potential and accelerations of the system}

\begin{figure}
\includegraphics[angle=0, width=80mm]{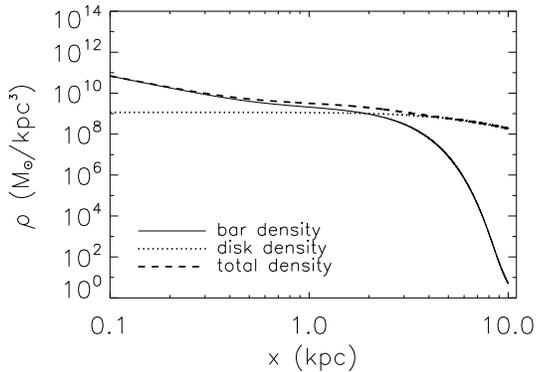}
\caption{Density distribution along the major ($x$-) axis. The solid,
dotted and dashed lines represent the bar, disk and total density of
the model, respectively.} \label{den}
\end{figure}

Figure~\ref{den} shows the density distribution for the models along the major axis ($x$-axis). It is obvious that the bar dominates the mass distribution of the
system in the inner 3\,kpc.

In order to solve the Poisson's equation, we follow
\cite{1992ApJ...386..375H} and \cite{1996MNRAS.283..149Z}  to expand
the potential and density on a set of simple orthogonal basis of
potential-density pairs in the spherical coordinates
\begin{equation}
\Phi_{\rm{bar}}=-\frac{GM_{\rm{bar}}}{r_{\rm
s}}\sum_{n,l,m}A_{nlm}\Phi_{nlm},
\end{equation}
where
\begin{equation}
\Phi_{nlm}=\frac{s^l}{(1+s)^{2l+1}}G_n^{(2l+3/2)}\bigg(\frac{s-1}{s+1}\bigg)P_{lm}(\cos\theta) \cos(m\phi),
\end{equation}
\begin{equation}
s=\frac{r}{r_{\rm s}}, r_{\rm s}=1\ \rm{kpc}
\end{equation}
and $G_n^{(2l+3/2)}(\xi)$ is the Gegenbauer polynomial of $\xi$. The
expansion coefficient $A_{nlm}$ can be calculated by
\begin{eqnarray}
A_{nlm}&=&\frac{1}{I_{nl}}\int_{-1}^{1}\frac{2}{(1-\xi)^2}d\xi\int_{-1}^{1}dX\int_0^{2\pi}d\phi
\rho(r,\theta,\phi)\nonumber\\
&&\times\bigg[RR(r)P_{lm}(X)\cos(m\phi)\bigg], \nonumber\\
\end{eqnarray}
where $\xi=\frac{r-1}{r+1}$, $X=\cos\theta$,
\begin{equation}
RR(r)=\frac{s^l}{(1+s)^{2l+1}}G_n^{(2l+3/2)}(\xi) r^2,
\end{equation}
\begin{eqnarray}
I_{nl}&=&K_{nl}\frac{1}{2^{8l+6}}\frac{\Gamma(n+4l+3)}{n!(n+2l+3/2)[\Gamma(2l+3/2)]^2}(1+\delta_{m0})\pi\nonumber\\
&&\times\frac{2}{2l+1}\frac{(l+m)!}{(l-m)!},
\end{eqnarray}
\begin{equation}
K_{nl}=\frac{1}{2}n(n+4l+3)+(l+1)(2l+1)
\end{equation}
and $\delta_{m0}$ is the Dirac Delta function which is defined as
$\delta_{m0}=1$ for $m=0$ and $\delta_{m0}=0$ for $m\neq0$.

\begin{figure}
\includegraphics[angle=0, width=80mm]{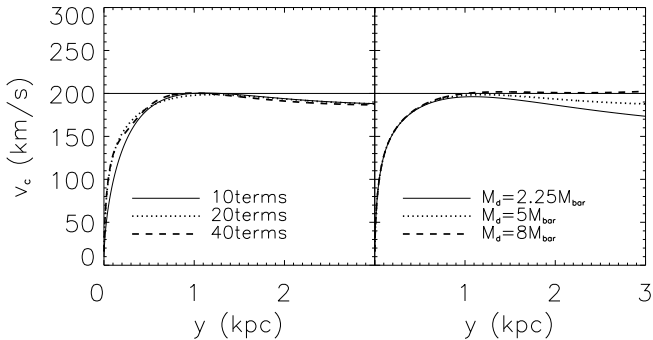}
\caption{Circular velocity along the intermediate ($y$-) axis of
the bar. \emph{Left}: The solid, dotted and dashed lines represent
the results from 10, 20 and 40 terms of the Hernquist-Ostriker
expansions, respectively. The disk mass is
$\rm{M_d=1.0\times10^{11}M_{\odot}}$. The flat horizontal line
indicates an amplitude of $200\kms$. \emph{Right}: The solid, dotted
and dashed lines represent results of $M_{\rm d}=2.25$, 5 and
8$M_{\rm bar}$, respectively. Twenty terms of Hernquist-Ostriker
expansions are used. } \label{vc}
\end{figure}

\begin{figure}
\begin{center}
\includegraphics[height=0.35\textwidth]{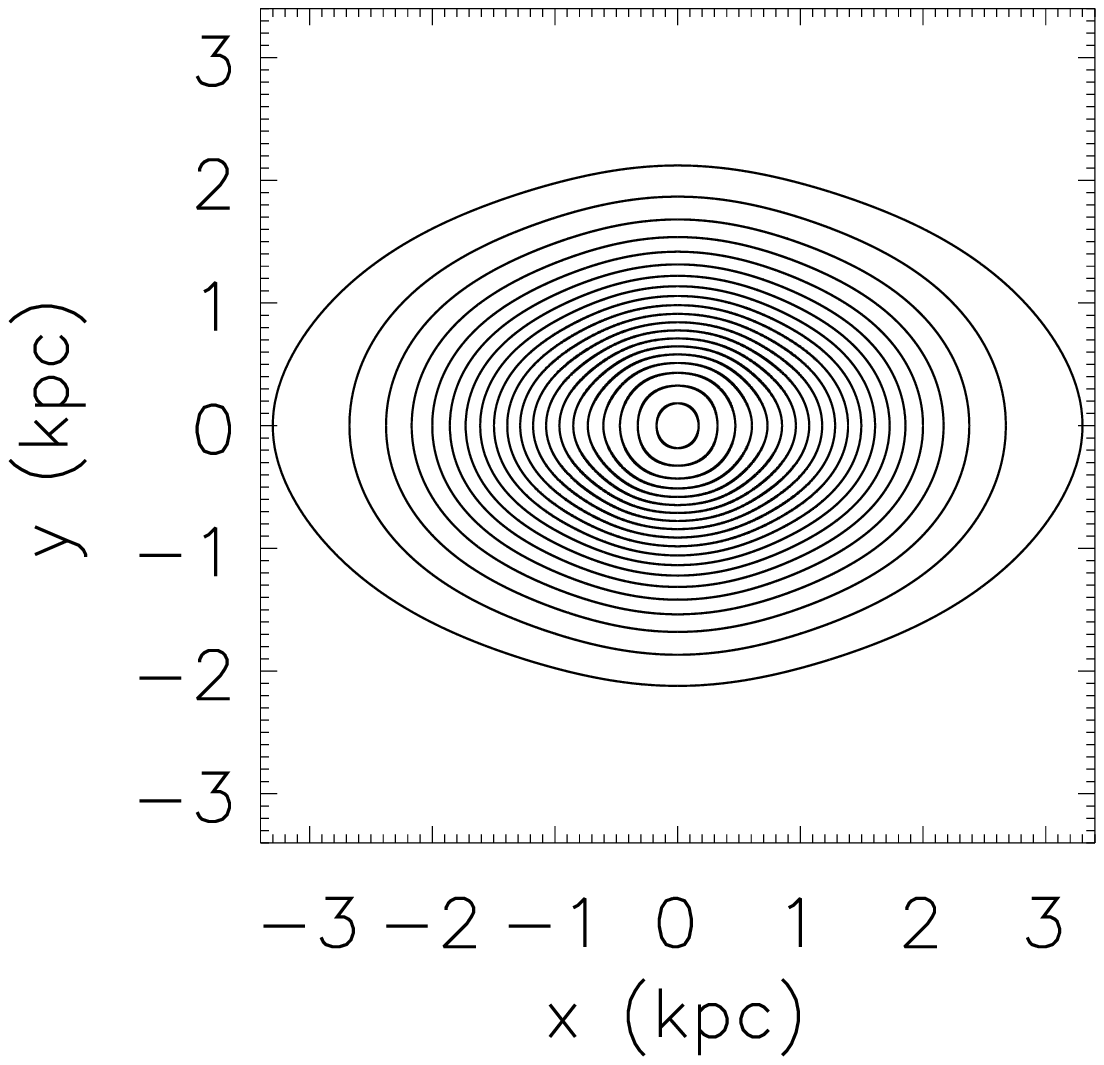}
\includegraphics[height=0.35\textwidth]{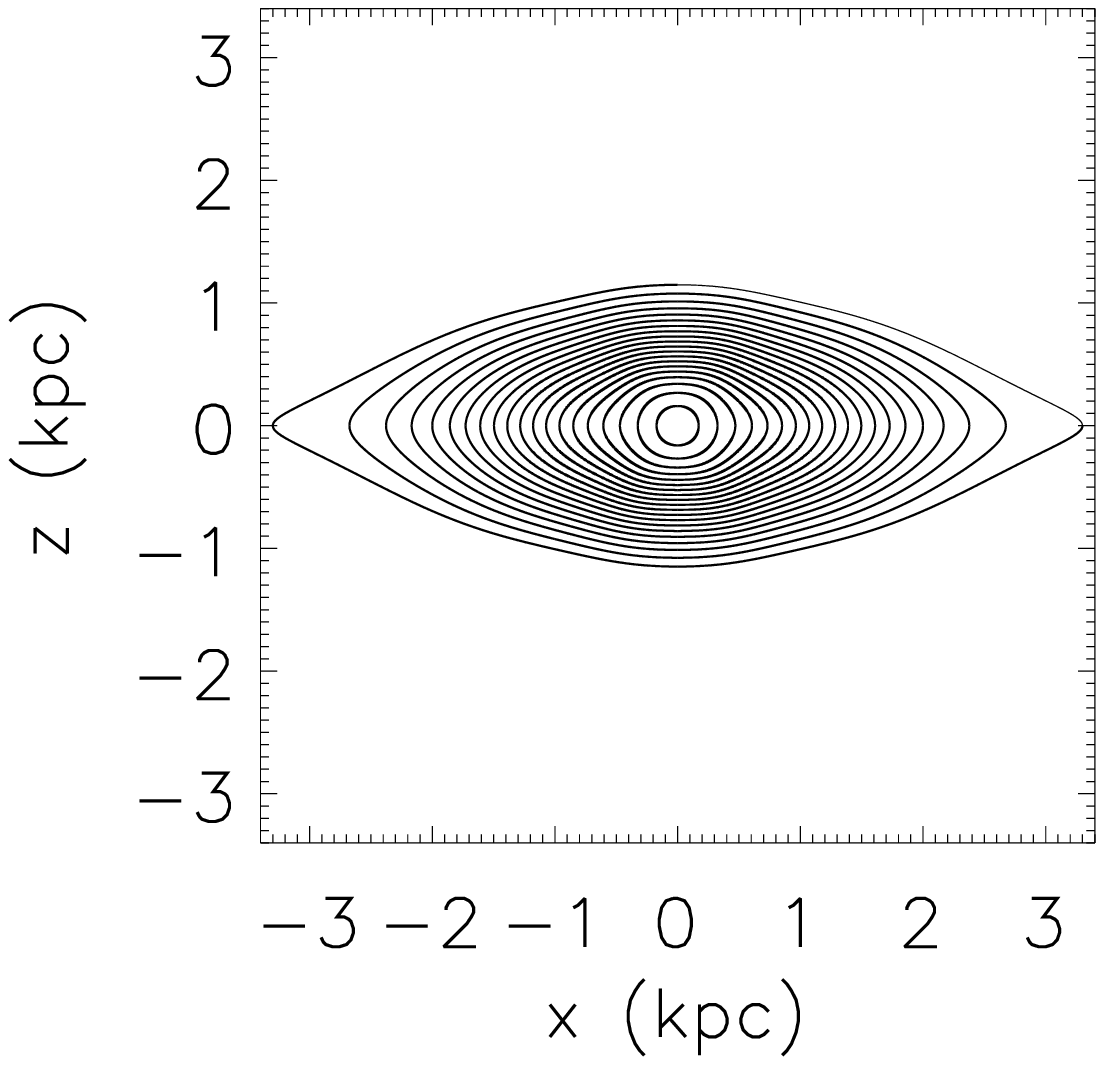}
\caption{\label{fig:iso} {Top}: Effective-potential contours for 20
 shells with equal mass in the $x-y$ plane. {Bottom}:
Effective-potential contours in the $x-z$ plane. The value of
effective potentials in the most inner and outer surfaces are
$-1.81\times10^{5}$ and $-1.08\times10^{5} ({\rm km\,s^{-1}})^2$,
respectively.
}\label{iso_pot}
\end{center}
\end{figure}

Each expansion coefficient is determined by the three quantum
numbers \emph{n}, \emph{l}, and\emph{ m}. For a triaxial model, only
the even quantum number terms are non-zero. The circular velocity
profile is shown in Figure ~\ref{vc} along the intermediate axis
($y$-axis). As can be seen, there is little difference between
models with 20 and 40 terms (left panel). So in this paper we will
adopt twenty expansion coefficients in the orbit integration to save
CPU time. They are listed in Table~\ref{tab:cnlm}. The right panel
of Figure~\ref{vc} shows the dependence of the circular velocity on
the disk mass. Clearly, the model with the massive disk mass has the
large circular velocity beyond $1\ \rm{kpc}$. Figure ~\ref{iso_pot}
shows the contour of effective potentials of the model in the $x-y$
plane (top panel) and $x-z$ planes (bottom panel).

\begin{table}
\caption{Twenty Hernquist-Ostriker expansion coefficients $A_{nlm}
$.} \label{tab:cnlm}

\begin{center}
\begin{tabular}{llllllllllllll}\hline
 n& l & m & $A_{nlm}$\\

 \hline\hline
 0  &0    &0     &1.509      \\
 1  &0    &0     &-0.086     \\
 2  &0    &0     &-0.033     \\
 3  &0    &0     &-0.020     \\
 0  &2    &0     &-2.606     \\
 1  &2    &0     &-0.221     \\
 2  &2    &0     &-0.001     \\
 0  &2    &2     &0.665      \\
 1  &2    &2     &0.129      \\
 2  &2    &2     &0.006      \\
 0  &4    &0     &6.406      \\
 1  &4    &0     &1.295      \\
 0  &4    &2     &-0.660     \\
 1  &4    &2     &-0.140     \\
 0  &4    &4     &0.044      \\
 1  &4    &4     &-0.012     \\
 0  &6    &0     &-5.859     \\
 0  &6    &2     &0.984      \\
 0  &6    &4     &-0.030     \\
 0  &6    &6     &0.001      \\

  \hline
 \end{tabular}
\end{center}

\end{table}

Since we have the potential for the systems, the accelerations can
be easily calculated from the potential; we do not give
detailed expressions for the acceleration here.

\section{model construction}
\subsection{Orbit-superposition technology}

Since \cite{1979ApJ...232..236S} pioneered the orbit-superposition
method to construct self-consistent models for three-dimensional
mass distribution, it has been widely
applied in dynamical modelling
\citep[e.g.][]{1997ApJ...488..702R,1998ApJ...493..613V,2005MNRAS.363..937B,2006A&A...445..513V,2007ApJ...666..165C,2008MNRAS.385..647V,2008ApJ...677.1033W,2009MNRAS.396..109W}.
The key point of this method is to construct an orbit library which
is sufficiently comprehensive in order to reproduce the available observations.

Specifically, let $N_{\rm o}$ be the total number of orbits and
$N_{\rm c}$ be the number of spatial cells. For each orbit $j$, we
count the fraction time $O_{ij}$ and projected quantities $P_{ij}$
that they spend in each cell $i$. The fraction time of $O_{ij}$ of
each orbit is obtained as follows. Every orbit is integrated for one
Hubble time ($t_H$), and $N=10,000$ output (position and velocity)
are stored at constant time interval  for each each orbit. If the
orbit $j$ crosses the cell $i$ one time, we will increase the number
$N_{ij}$ by 1. Then, the fraction $O_{ij}$ is determined by
$O_{ij}=N_{ij}/N$. The orbit weight $W_j$ for each orbit $j$ is then
determined by the following equations:
\begin{equation}\label{wei1}
\mu_i=\frac{\sum_{j=1}^{N_{\rm o}}W_jO_{ij}P_{ij}}{\sum_{j=1}^{\rm{N_{\rm o}}}W_jO_{ij}},~~~~~~~~
i=1,.....,N_{\rm c}
\end{equation}
where $\mu_i$ can be the volume density, surface density, or moments
of the velocity distribution in each cell $i$. If $\mu_i$ is the
mass or the velocity, then equation~(\ref{wei1}) is same as equation
(2.4) or (2.5) of ~\cite{1984A&A...141..171P}. Following ZH96, we
divide the first octant into 1000 cells with similar masses. Due to
symmetry of the model, the other octants are ``reflected'' to the
first octant. In each (x-, y-, and z-) direction, the system is
divided into 10 bins. Each cell is a small box with $dx=0.25$\,kpc,
$dy=0.15$\,kpc and $dz=0.10$ kpc.  The central cell covers the box
with $x=[-0.25,0.25]$, $y=[-0.15,0.15]$ and $z=[-0.1,0.1]$.

More practically, equation~\ref{wei1} can be written as a set of
linear equations
\begin{equation}\label{wei1b}
 \sum_{i=1}^{\rm{N_{\rm c}}}(\mu_i-P_{ij})O_{ij}W_j=0
\end{equation}

We adopt the non-negative least squares (NNLS) method
\citep{1984A&A...141..171P}  to solve $W_j$, i.e., the following
$\chi_w^2$
\begin{equation}\label{wei2}
\chi_w^2=\bigg|\sum_{i=1}^{\rm{N_{\rm
c}}}(\mu_i-P_{ij})O_{ij}W_j\bigg|^2
\end{equation}
is minimized with respect to $W_j$ ($j=1, ......, N_{\rm o}$) to
obtain the values of $W_j\ge 0$. It is obvious that the NNLS fit
will find a unique solution if the number of orbits is smaller than
the number of constraints. However, such a model may not be
self-consistent. A more meaningful result with the NNLS method
should use a large number of smooth orbits well sampled in the phase
space. In this case, many exact solutions with $\chi^2=0$ are
possible. The NNLS method will select one of the possible solutions.

Our density model is smooth, therefore, we expect that the
phase-space density from the reconstructed self-consistent model is
also smooth. We employ a simple smoothing procedure to fit the data.
We require simply that the orbits with adjacent initial conditions
have nearly the  same weight (\citealt{1996ApJ...460..136M},
hereafter MF96). In this approach, Equation ~\ref{wei2} becomes,
\begin{equation}\label{wei3}
\chi_w^2=\bigg|\sum_{i=1}^{\rm{N_{\rm
c}}}(\mu_i-P_{ij})O_{ij}W_j\bigg|^2+\lambda\sum_{j=1}^{N_{\rm
o}}W_j^2
\end{equation}
where $\lambda$ is a smoothing parameter and $\lambda=N_{\rm
o}^{-2}$.

\begin{figure}
\includegraphics[angle=0, width=80mm]{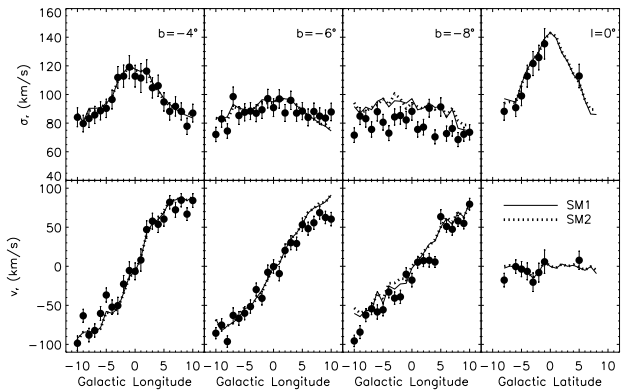}
\caption{Radial velocity and velocity dispersion along four strips
with $b=-4^{\circ}$, $b=-6^{\circ}$, $b=-8^{\circ}$ and
$l=0^{\circ}$. The solid (SM1) and dotted (SM2) lines indicate the
results from our models using the smoothing method as shown in M96
and ZH96, respectively. The filled circles with error bars are data
from BRAVA \citep{2012AJ....143...57K}. } \label{vr_smooth}
\end{figure}

MF96's method strongly depends on the size of the cells. In order to
assess whether the smoothing method affects our results, we also
used another smoothing method, also adopted by ZH96. The key point
of this smoothing method is that orbits with similar integrals of
motion should have similar orbit weights. In a rotational bar
system, only Jacobi's integral $E_J$ is an integral of motion
\citep{1987gady.book.....B}, here we take the time averaged
quantities $\langle L_z \rangle $ and $\langle L_x^2 \rangle$ as the
effective integrals, where $L_z$ and $L_x$ are an orbit's
instantaneous angular momentum components along the minor and major
axes, respectively. Orbits with similar $E_J$, $\langle L_z \rangle
$ and $\langle L_x^2 \rangle$ have similar orbit weights. We have
checked the results from these two smoothing methods and found no
significant difference (See Figure~\ref{vr_smooth}) in their
abilities to fit the BRAVA data. The only difference is that we find
the MF96's method yields more orbits with non-zero weights than
those of ZH96. From now on, we only present results using the
smoothing method in M96.

\subsection{Constructing the orbit library}
\label{sec:library}

In a rotational triaxial system, as mentioned before only Jacobi's
energy is an integral of motion. Early studies have shown that most
orbits are chaotic \citep[see also section
\ref{sec:orbits}]{2007MNRAS.381..757V, 2011arXiv1102.1157M} and the
typical regular orbits are $x_1$-type
\citep{1987gady.book.....B,1994PhDT.........5Z,1994AJ....108.2154Z}
in a bar model. We do not know the explicit phase space distribution
$f(x, y, z, v_x, v_y, v_z)$ for the chaotic orbits because they lack
integrals of motion in a rotating bar potential, we will thus
consider two different methods to generate the initial conditions of
orbits. The common point of these methods is that the bar model is
divided into 20 shells with nearly equal mass spatially along the
$x$-axis by using the Monte Carlo integration (see Fig.
\ref{iso_pot}).

The first method (IC1) is similar to the one adopted in
ZH96. Here we give a brief description of its main ingredients, and
refer the reader to Appendix B in ZH96 for more details. The
orbits are launched in close pairs perpendicular to the $xz-$, $yz$-
or $xy-$ symmetry plane or the $x$, $y$ axis. The initial position of
each orbit in each shell has the same effective potential, which is
defined as that in the bounded surface of two
close shells. The initial velocity is tangential and less than the
local circular velocity. Only 1000 initial conditions of orbits were
generated in ZH96, here we extend the number of orbits to
$\sim20,000$.

The second method (IC2) is similar to the one used in
\cite{2000MNRAS.314..433H}.
The initial conditions are generated with known distribution functions
$f=\sum_{i=1}^{3}C_i\rho(x, y, z)h_i$, where $C_i, i=1, 2, 3$ are three
normalizing constants, $\rho$ is the density distribution of the
system. The three functions $h_{i=1,2,3}$ are defined as
\begin{equation}
h_3(v_x,v_y,v_z)=\frac{1}{(2\pi)^{3/2}\sigma_x\sigma_y\sigma_z}\exp\bigg(-\frac{v_x^2}{2\sigma_x^2}-\frac{v_y^2}{2\sigma_y^2}-\frac{v_z^2}{2\sigma_z^2}\bigg)
\end{equation}
and
\begin{eqnarray}
&&h_{1,2}(v_R,v_{\phi},v_z)=\frac{1}{(2\pi)^{3/2}\sigma_R\sigma_{\phi}\sigma_z}\nonumber\\
&&\times\exp\bigg[-\frac{v_R^2}{2\sigma_R^2}-\frac{(v_{\phi}\mp
v_{c1})^2}{2\sigma_{\phi}^2}-\frac{v_z^2}{2\sigma_z^2}\bigg]
\end{eqnarray}
where $v_{c1}$ is defined by
\begin{equation}
v_{c1}(R)=250 \kms \bigg[1+\bigg(\frac{0.1\ \rm{kpc}}{R}\bigg)^{0.2}
\bigg]^{-1}.
\end{equation}
 For the velocity dispersion parameters, we adopt the same values as listed in
 Table 2 of \cite{2000MNRAS.314..433H}. We select 50,000 initial
 conditions using this method.

\begin{figure}
\includegraphics[angle=0, width=80mm]{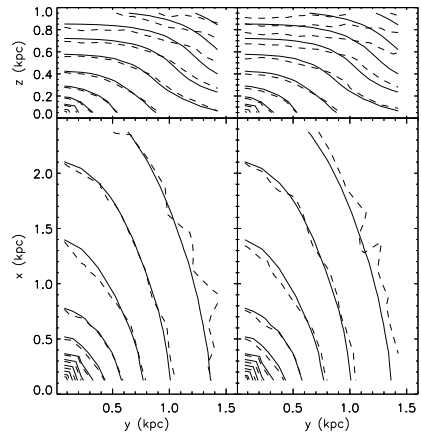}
\caption{Density contours for the input model (solid lines) and from
orbits (dashed lines) in the $x-y$ and $y-z$ planes. The left and
right panels represent the results from IC1 and IC2, respectively
(see \S\ref{sec:library}).} \label{iso_3den_hs_bn}
\end{figure}

\begin{figure}
\includegraphics[angle=0, width=80mm]{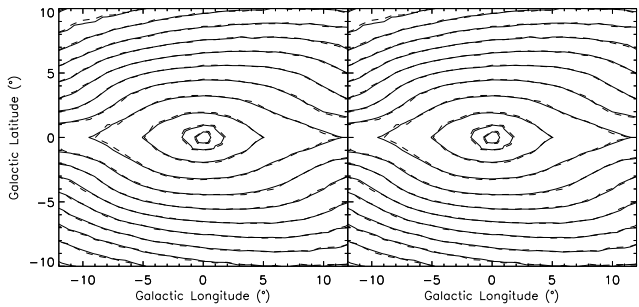}
\caption{Similar to Figure~\ref{iso_3den_hs_bn}, but for the
projected surface density.} \label{iso_2den_hs_bn}
\end{figure}

In Figures~\ref{iso_3den_hs_bn} and ~\ref{iso_2den_hs_bn}, we
present the volume density and the projected density contours for
model 25 (see Table 1), respectively. In each figure, the solid and
dashed lines represent the results from the input and orbit models,
respectively. The left and right panels represent the results from
the two methods (IC1 and IC2), respectively. It is seen that the
orbit from both methods can reproduce the density distribution and
projected density distribution well. The difference between the
reconstructed and input densities is small, which
indicates that our model is self-consistent.

\begin{figure}
\includegraphics[angle=0, width=80mm]{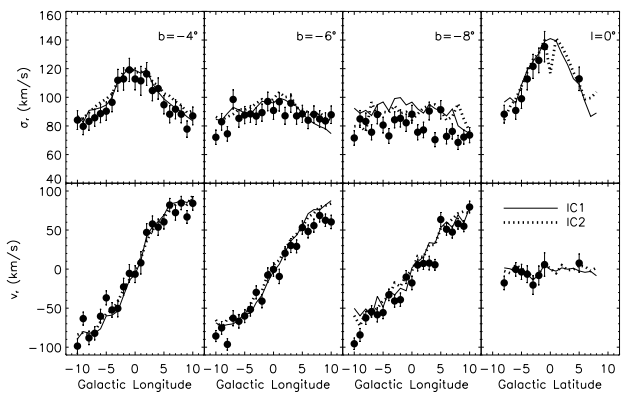}
\caption{{\sl BRAVA} data versus model as in Figure~\ref{vr_smooth}.
The solid (IC1) and dashed (IC2) lines indicate the results from our
model, while the filled circles with error bars are data from
BRAVA.} \label{vr_IC}
\end{figure}

Figure \ref{vr_IC} further compares the velocity and velocity dispersion
from the model with those from the BRAVA data. There is no significant
difference between the results from these two initial conditions, thus
from now on, we only present results from the first method (IC1).

\begin{figure}
\begin{center}
\includegraphics[height=0.35\textwidth]{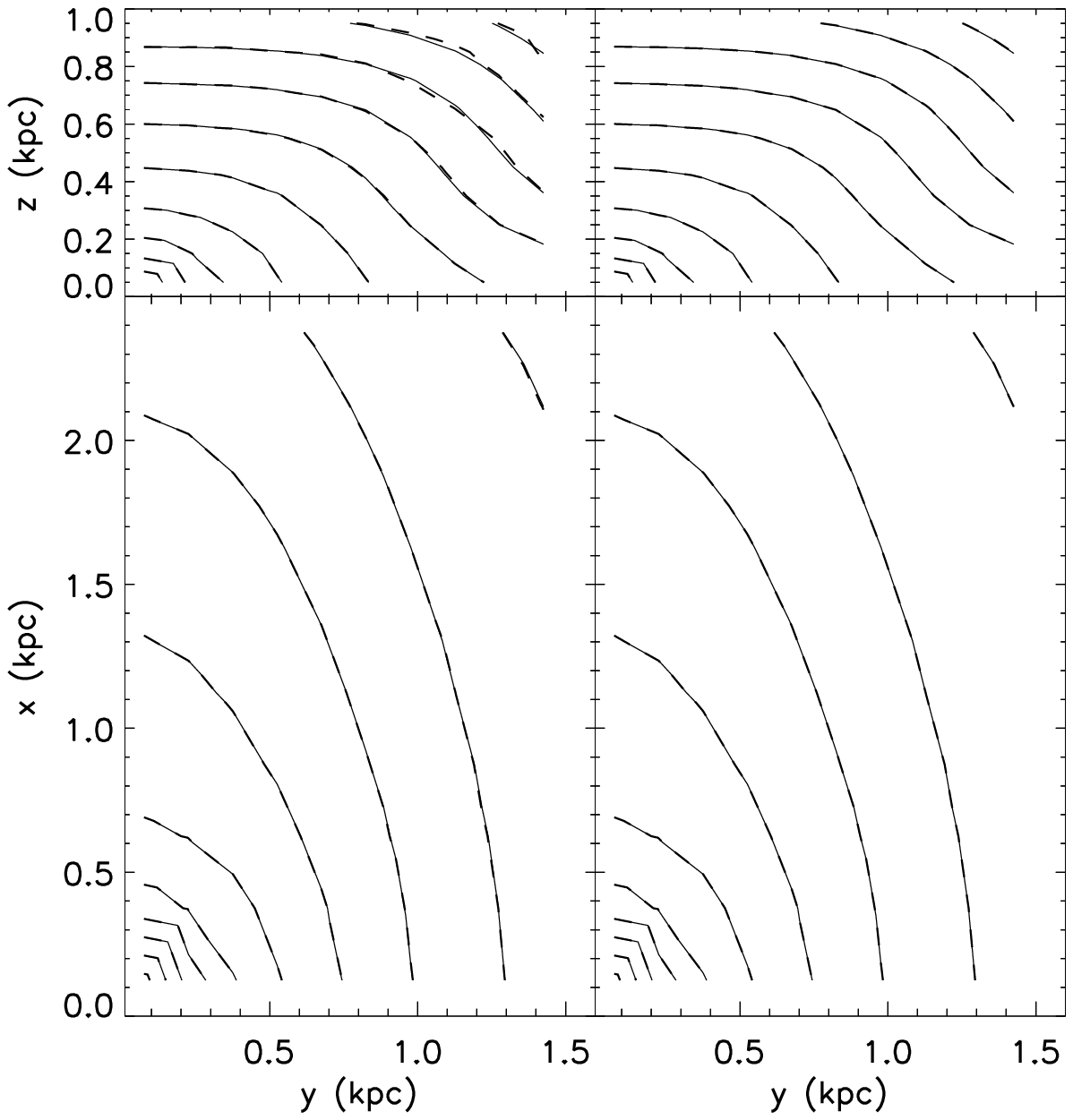}
\includegraphics[height=0.35\textwidth]{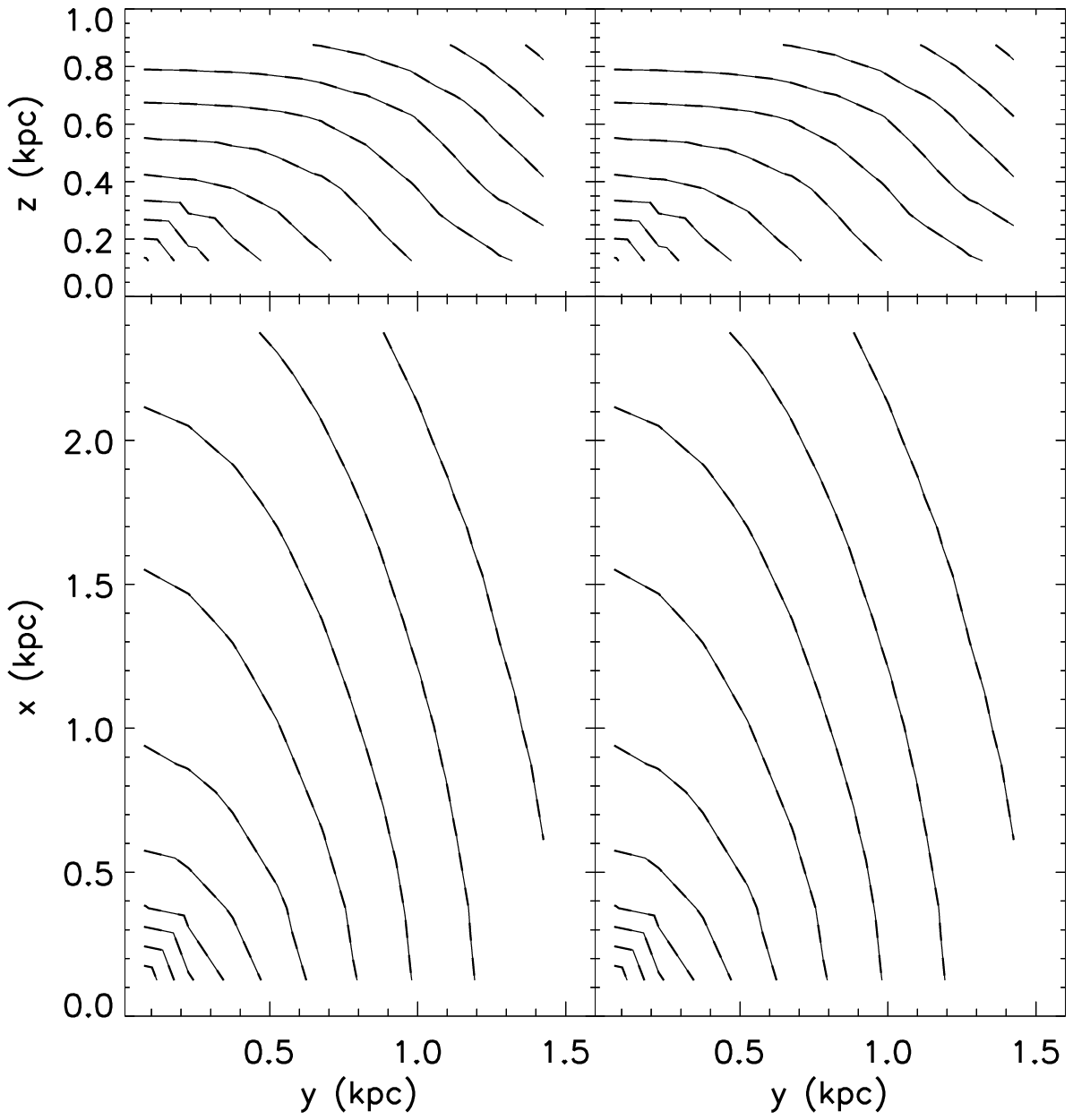}
\caption{Mass distribution for the input model (solid lines) and
from orbits (dashed lines) in the $x-y$ and $y-z$ planes. The left
and right panels represent the results from the first and second
half Hubble time. Top: Results for using 1000 cells. Bottom: Results
for using 400 cells.}\label{oij}
\end{center}
\end{figure}

As pointed out by \cite{1984A&A...141..171P}, a reasonable orbit
integration time may be determined by the fluctuation of the
$O_{ij}$ between two successive halves of the integration time. The
$O_{ij}$ actually reflects the orbit densities and a superposition
of $O_{ij}$ reflects the system density. For regular orbits,
$O_{ij}$'s can reach stable values in a relatively short time.
However, $O_{ij}$'s for irregular orbits only can converge after a
very long integration time, at least 1000 Hubble times as suggested
by \cite{1984A&A...141..171P}. In practice, it will be too
time-consuming to integrate a large number of orbits for such a long
time. We have compared the fluctuation of the $O_{ij}$'s between two
successive halves in one Hubble time with those in ten Hubble times
for irregular orbits, there is no clear improvement in the
convergence. In this paper, the orbits are integrated for one Hubble
time unless stated otherwise. Although the typical fluctuation  of
$O_{ij}$'s is about $\sim 20\%$ between the first and second half,
the typical mass fluctuation in each cell is small (below $2\%$), as
can be seen from Figure~\ref{oij}. Figure~\ref{oij} shows the mass
distribution when the integration time is the first and second half
Hubble time respectively. We only consider the mass distribution in
Equation~(\ref{wei2}), and no smoothing is adopted. In order to
decrease the fluctuation, we decrease the spatial resolution of
cells, reducing the cell number from 1000 to 400 by merging every
two and half adjacent cells into one along the z-direction. The
typical fluctuation of the $O_{ij}$ between the first and second
half Hubble time is reduced to $\sim10\%$ and the mass fluctuation
is $\sim$ $0.05\%$.

Since there are many parameters in the bar model, unless stated
otherwise, we adopt $M_{\rm bar}={2.0\times10^{10}M_{\odot}}$,
$M_{\rm d}=8M_{\rm bar}$, $\Omega_{\rm p}=60\ \rm{km\ s^{-1}\
kpc^{-1}}$ and bar angle $\theta_{\rm bar}=13.4^{\circ}$. These
parameters are the same as those used in ZH96 (model 25 in Table
\ref{tab:model}) for ease of comparisons with ZH96.

\section{results}

\subsection{Model constraints}

The key point is to solve the orbit weights from equation
~\ref{wei2}. To do this, the volume density, projected density,
radial velocity and velocity dispersion along four windows
$b=-4^{\circ}$, $b=-6^{\circ}$, $b=-8^{\circ}$ and $l=0^{\circ}$
(see Figure~\ref{vr_smooth}) are used as constraints to solve the
weight of each orbit. The volume density and projected density are
obtained directly from the density distribution of the bar and
Miyamoto-Nagai disk model. Since our aim is to construct a
self-consistent bar model, the volume density is fitted only inner
3kpc around the Galactic center.

\begin{figure}
\includegraphics[angle=0, width=80mm]{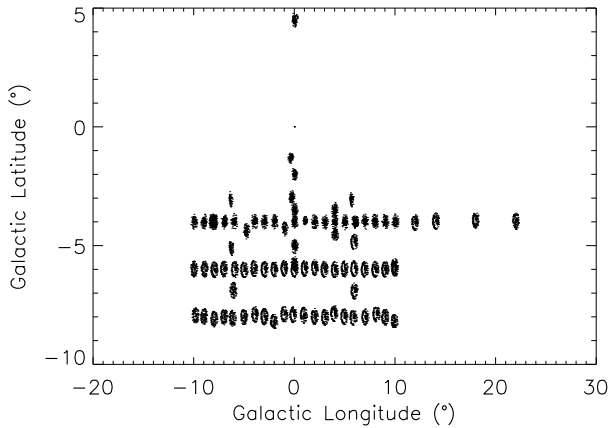}
\caption{Observed fields by the BRAVA survey from 2005 to 2008 (also
see Fig. 1 in Kunder et al. (2012)).} \label{brava}
\end{figure}

The kinematic constraints (radial velocity and dispersion) are
from the BRAVA survey, conducted from 2005 to 2008 in 8746 fields, which are shown
in Figure \ref{brava}. It is seen that most data are located in the range
$l=[-12^{\circ},12^{\circ}]$ and $b=[-10^{\circ},10^{\circ}]$, where
$l$ is the Galactic longitude and $b$ is the Galactic latitude. In
this paper, we fit the projected density, the radial velocity and
velocity dispersion in the range $l=[-12^{\circ},12^{\circ}]$,
$b=[-10^{\circ},10^{\circ}]$.

\subsection{Dependence on model parameters}
\label{sec:dependence}

In this subsection, we vary the model parameters around the ZH96 model (model 25 in Table \ref{tab:model}) to see the trends. We will explore the parameter space more systematically in section \ref{sec:best-fit}.

\subsubsection{Bar angle}

From the COBE map, we know that there is a clear offset between the
major axis of the bar and the line-of-sight to the Galactic center.
 However, its value is not accurately known.
From COBE observations, \cite{1995ApJ...445..716D} found the value
of bar angle is $20^{\circ}\pm10^{\circ}$. Using the same COBE map,
\cite{1994PhDT.........5Z} obtained a bar angle $13.4^{\circ}$ while
\cite{1997MNRAS.288..365B} found $20^{\circ}$.
\cite{2000ApJ...541..734A} found this value to be $~15^{\circ}$.
Recently, the 6.7-GHz methanol masers showed a $45^{\circ}$
orientation of bar angle \citep{2011arXiv1103.3913G}. No consensus appears to be emerging among the recent observations concerning the bar angle.

\begin{figure}
\includegraphics[angle=0, width=80mm]{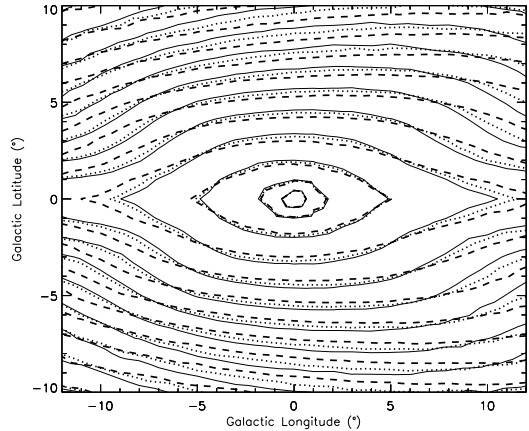}
\caption{Projected density of the input bar model for different bar
angles. The solid, dotted, and dashed lines represent the results
for the bar angle $13.4^{\circ}$, $20^{\circ}$, and $30^{\circ}$,
respectively.} \label{iso_2den_angle}
\end{figure}

In Figure~\ref{iso_2den_angle}, we show the projected density maps
of the input bar model for different bar angle. It is clear that the
isodensity maps become more sharp-edged with an increasing bar
angle, which means that the bar angle can be determined if high
quality surface brightness data are available. In this paper, our
bar model is from ZH96, which attempts to match COBE observations.
We can vary the bar angle from $13.4^{\circ}$ to $20^{\circ}$ or
even $30^{\circ}$ within the error bar of the model. However, a
large bar angle ($40^{\circ}$) would require us to refit  data in
terms of other bar parameters (lengths and axial ratios,
\citealt{1996MNRAS.282..175Z}). Therefore, we restrict ourselves to
three different bar angles $13.4^{\circ}$, $20^{\circ}$ and
$30^{\circ}$.

\begin{figure}
\includegraphics[angle=0, width=80mm]{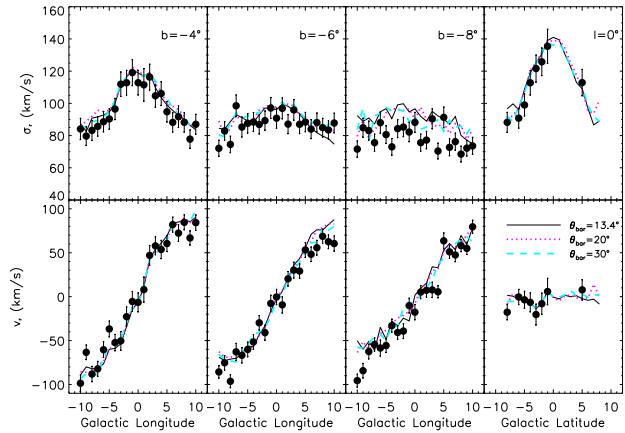}
\caption{Radial velocity and velocity dispersion along the bulge
major axis for different bar angles. The pattern speed is
$\Omega_{\rm p}=60\ \rm{km\, s^{-1}\, kpc^{-1}}$ and $M_{\rm
d}=8M_{\rm bar}$, where $M_{\rm bar}=2 \times 10^{10} M_{\odot}$.}
\label{vr_angle}
\end{figure}

Figure~\ref{vr_angle} compares the radial velocity and velocity
dispersion from the
orbit projection with ones from the BRAVA data for different angles.
There is a small difference between the results from different bar
angles. However, it is clear that kinematics alone constrain the bar
angle poorly.

\subsubsection{Pattern speed}

The pattern speed of the Galactic bar has been estimated from
different methods and are somewhat uncertain. \cite{2002MNRAS.334..355D} used the
Tremaine-Weinberg continuity method to the OH/IR stars and obtained
a value $\Omega_{\rm p}=(59\pm5\pm10)\,\rm{km s^{-1}\ kpc^{-1}}$.
\cite{1999MNRAS.304..512E} obtained $\Omega_{\rm p}\approx60\, \rm{km
s^{-1} kpc^{-1}}$ by comparing the gas flow in hydrodynamic
simulations with the velocity curve from HI and CO observations.
From the length of the bar
\citep{1997MNRAS.288..365B,2005ApJ...630L.149B,2007AA...465..825C},
the pattern speed was given in a wide range $\Omega_{\rm p}\sim (35-60)
$ $\rm{km s^{-1} kpc^{-1}}$ \citep{2010arXiv1003.2489G}.

\begin{figure}
\includegraphics[angle=0, width=80mm]{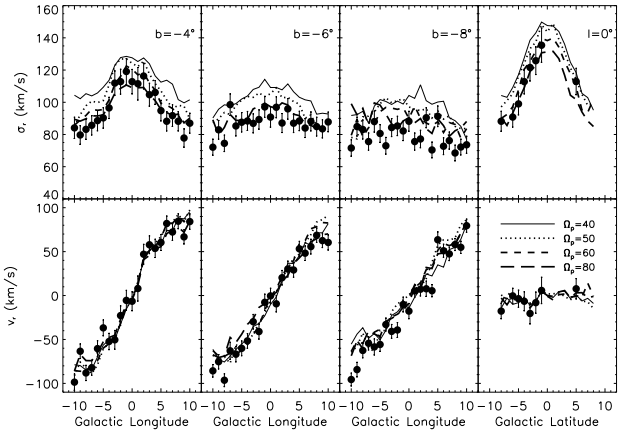}
\caption{Similar to Figure~\ref{vr_angle}, but for different pattern
speeds of the bar. The solid, dotted dashed and long dashed lines
represent $\Omega_{\rm p}=40$, 50, 60 and 80 $\rm{km\ s^{-1}\
kpc^{-1}}$, respectively. $M_{\rm d}=8M_{\rm bar}$, $\theta_{\rm
bar}=20^{\circ}$.} \label{vr_omp}
\end{figure}

We consider four pattern speeds $\Omega_{\rm p}=40$, 50, 60 and 80
$\rm{km s^{-1} kpc^{-1}}$. In Figure~\ref{vr_omp}, we present the
velocity and velocity dispersion distributions for different pattern
speeds. Obviously, the velocity dispersion profile strongly depends
on the pattern speed. The predicted velocity dispersion
of a model is inversely correlated with its pattern speed.

\subsubsection{Disc mass}

\begin{figure}
\includegraphics[angle=0, width=80mm]{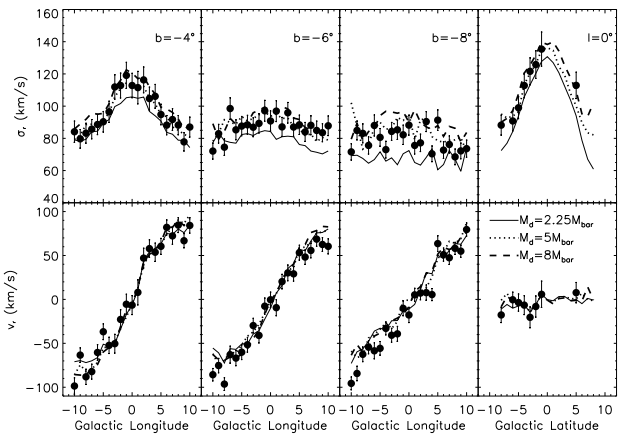}
\caption{Similar to Figure~\ref{vr_angle}, but for different values
of the disk mass. The solid, dotted and dashed curves represent
$M_{\rm d}=2.25$, 5 and 8 $M_{\rm bar}$, respectively. $\Omega_{\rm
p}=60\ \rm{km\ s^{-1}\ kpc^{-1}}$, $\theta_{\rm bar}=20^{\circ}$. }
\label{vr_disk}
\end{figure}

The disk mass is another parameter which is not accurately known. We
consider three different values of the disk mass.
Figure~\ref{vr_disk} shows the dependence of the velocity and
velocity dispersion on the disk mass. As can been seen, the velocity
dispersion profile from the model strongly depends on the value of
the disk mass: as expected, a less massive disk induces a lower velocity
dispersion than a more massive one.

\subsection{Best-fit model}
\label{sec:best-fit}

As mentioned in section \ref{sec:dependence},
 the kinematics from the model depend on
the bar angle, pattern speed and disk mass. In principle, we can
divide the parameter space of the bar angle, patter speed and disk mass
into many cells. In each parameter cell, we can run
the orbit-superposition technique and use the $\chi^2$ fit to
find the best-fit parameters. However, numerical calculation is
expensive. Here, we calculate 36 models with different parameters
(bar angle, pattern speed and disk mass). The $\chi^2$ is defined as
\begin{equation}
\chi^2=\sum_{i=1}^{\rm{N_d}}\frac{(y_{\rm obs}-y_{\rm
model})^2}{\sigma_{\rm obs}^2},
\end{equation}
where $\rm{N_{\rm obs}}$ is the total number of observed data,
$y_{\rm obs}$ and $y_{\rm model}$ are the observed and model
kinematics, respectively.
Since the three-dimensional and projected densities are given by the
input model, we do not include them in the $\chi^2$ fitting, although we
do compare the predicted
distributions with data by eyes.

Table~\ref{fit} lists the value of $\chi^2$ for fitting velocity and
velocity dispersion along both the major and minor axis of the bar
for 36 models. It is seen that models 12, 13, 14, 15, 22, 23, 24, 33
have smaller values of $\chi^2$ than others. visual examination
indicates that model 23 is the best-fit model in both the
three-dimensional and projected density distributions. Therefore, we
choose it as the best-fit model for the BRAVA data.

\begin{figure}
\includegraphics[angle=0, width=80mm]{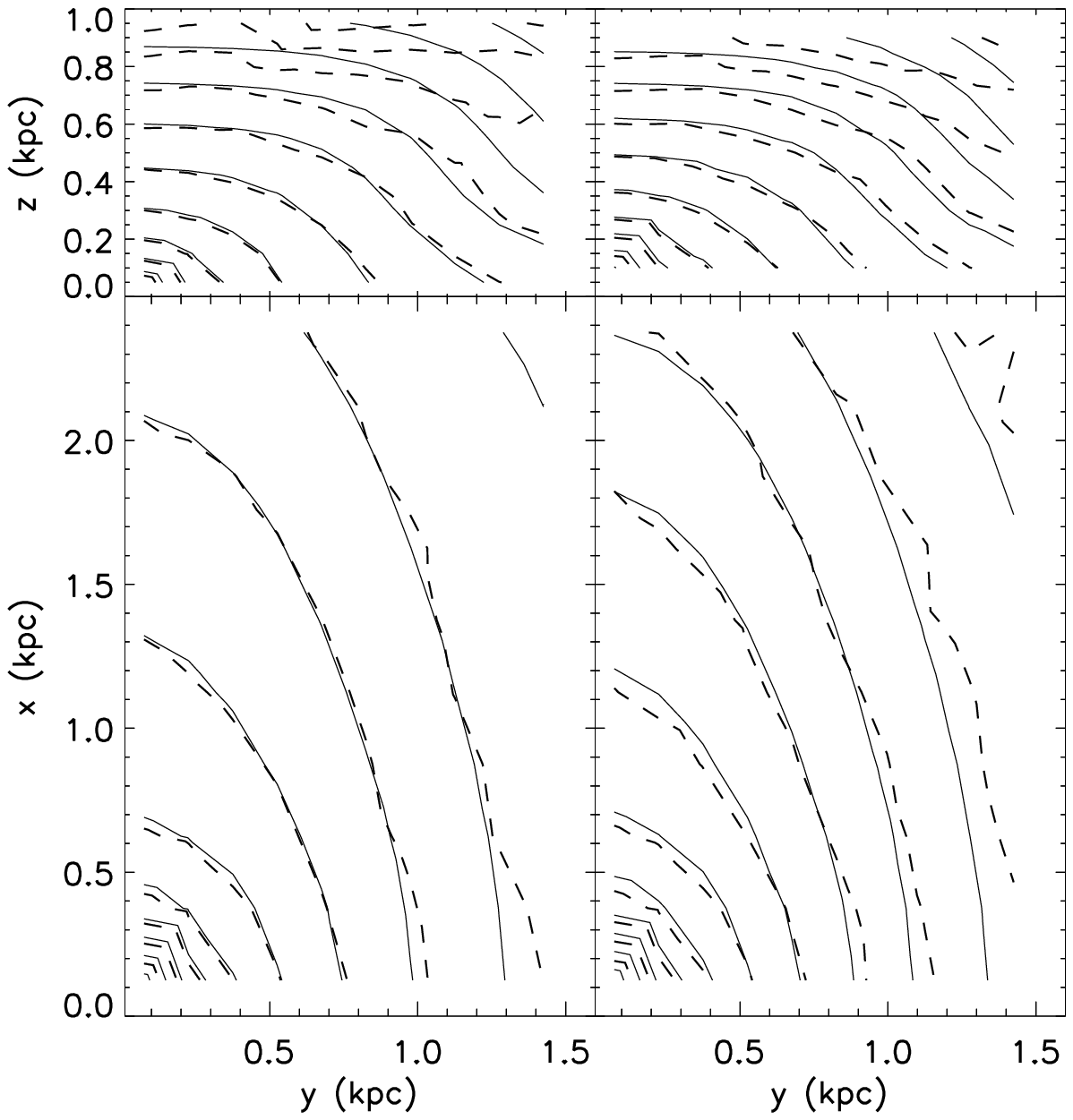}
\caption{Density contours from the input model (solid lines) and
from orbits (dashed lines) in the $x-y$ and $y-z$ planes for model
23. The left and right panes show the results by using 1000 and 400
mass cells, respectively. }\label{3d_23}
\end{figure}

\begin{figure}
\includegraphics[angle=0, width=80mm]{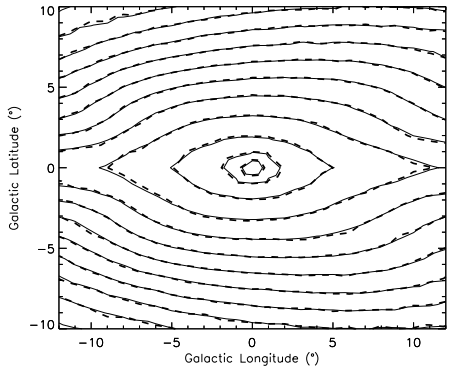}
\caption{Projected density contours from the input model (solid
lines) and from orbits (dashed lines) for model 23. 1000 cells are
used for solving the orbit weight. }\label{2d_23}
\end{figure}

\begin{figure}
\includegraphics[angle=0, width=80mm]{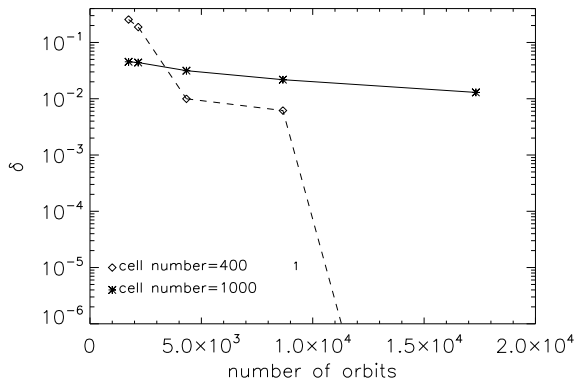}
\caption{Departure from self-consistency $\delta$ as a function of
the number of orbits. The star and diamond symbols represent the
fitting results by using 1000 and 400 mass cells, respectively.
}\label{delta}
\end{figure}

In Figures~\ref{3d_23} and ~\ref{2d_23}, we present the volume
density and the projected density contours, respectively. In each
figure, the solid and dashed lines represent the results from the
input and orbit models, respectively. It is seen that the orbit from
model 23 can reproduce the density distribution and projected
density distribution well. 
Moreover, we also use a parameter $\delta$ to describe the departure
from self-consistency for model 23, which is defined as (MF96):

\begin{equation}
\delta=\sqrt{\chi_w^2}/\bar{M},
\end{equation}
where $\bar{M}$ is the average mass in each cell, if the total mass
is 1, then $\bar{M}=1/N_c$. The $\chi_w^2$ is obtained from
equation~\ref{wei2} by only using the mass constraints and without
smoothing. Figure~\ref{delta} shows the departure from
self-consistency as a function of the number of orbits. It is noted
that departure parameter $\delta$ strongly depends on the number of
orbits and the cell number, $\delta$ is smaller than $10^{-6}$
(0.013) when 17,323 orbits are adopted in equation~\ref{wei2} by
using the 400 (1000) mass cells, which again shows that model 23 is
nearly self-consistent.

\begin{figure}
\includegraphics[angle=0, width=80mm]{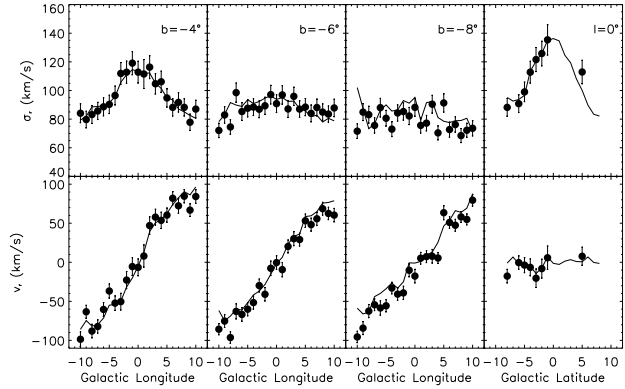}
\caption{Velocity and velocity dispersion distribution for model 23
($\Omega_{\rm p}=60\ \rm{km\ s^{-1}\ kpc^{-1}}$, disk mass
$\rm{M_d=1.0\times10^{11}M_{\odot}}$ and bar angle $\theta_{\rm
bar}=20^{\circ}$). The solid line is for the model while the filled
circles are data from BRAVA.} \label{vr_omp60}
\end{figure}

The velocity and velocity dispersion distributions of model 23 in
Figure~\ref{vr_omp60}. Note that model 23 can fit the BRAVA data
well with only a few outliers.

\subsubsection{Predicted proper motions}

Proper motions are not taken into consideration in solving
Equations~\ref{wei2} and \ref{wei3} because the absolute values of
proper motions can not be obtained from observations at present due
to the lack of absolute astrometry. We can still compare the
predicted proper motion dispersions with those observed.
Table~\ref{pm} shows the proper motions in some observed fields
together with the predictions from model 23. It is seen that the
proper motions along the latitude from model 23 in the Baade's and
Sagittarius's Window are in good agreement with those in
observations, while the latitudinal proper motion in the Plaut's
Window is lower than that in observations. The predicted proper
motions along the longitude in the three windows are greater than
observed. There are also more proper motion data in small fields
from OGLE \citep{2004MNRAS.348.1439S,
2007MNRAS.378.1165R,2008MNRAS.385..905R} and HST
\citep{2006MNRAS.370..435K}, we do not compare with the proper
motion in these field because the sky area in our model is
$1^{\circ}\times 1^{\circ}$, much larger than the observed field
size.

\subsubsection{Phase-space distribution and orbit families}
\label{sec:orbits}

\begin{figure}
\includegraphics[angle=0, width=80mm]{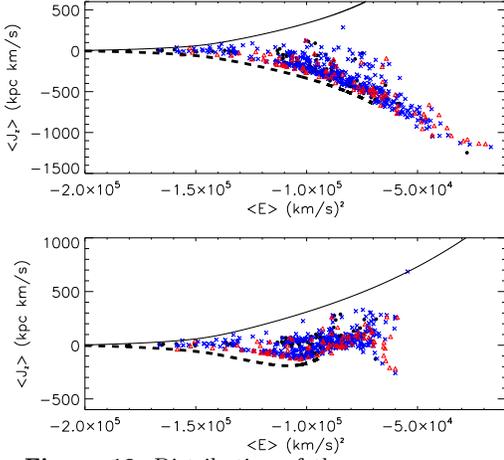}
\caption{Distribution of the average energy versus angular momentum
along the $z$-axis for orbits with non-zero weights. The upper and
lower panels show the results in the laboratory and rotational
frames, respectively. The filled circles, crosses and triangles
represent orbits with large, intermediate and small weights,
respectively. The solid and dashed lines represent the theoretical
retrograde and prograde motions from 0 to 3kpc, respectively. The
bar model is model 23 with $\Omega_{\rm p}=60\ \rm{km\ s^{-1}\
kpc^{-1}}$, $M_{\rm d}=5.0\times M_{\rm bar}$ and $\theta_{\rm
bar}=20^{\circ}$.} \label{E_Jz_omp60}
\end{figure}

The phase-space distribution and the orbit family are useful to help
us understand the model. Figure~\ref{E_Jz_omp60} shows the
distribution of average energy versus angular momentum along the
$z$-axis for non-zero weight orbits. The solid and dashed lines are
the theoretical distributions of the energy versus angular momentum
along the $z$-axis for retrograde and prograde motions from 0 to
3kpc, respectively. The energy and angular momentum along the
$z$-axis in the laboratory frame are defined as $E_{\rm
lab}(r)=\Phi(r)+ [\pm|V_{\rm c}|]^2/2$, $J_{\rm
z,lab}(r)=[\pm|V_{\rm c}|]\times r$. In the rotating frame, they are
$E_{\rm rot}(r)=\Phi(r)+ [\pm|V_{\rm c}| + |\Omega_{\rm p}|r]^2/2$,
and $J_{\rm z,rot}(r)= [\pm|V_{\rm c}| + |\Omega_{\rm p}|r] \times
r$. Here $r$ is the radius, $V_{\rm c}$ is the circular velocity and
$\Phi$ is the potential of the system. The `+' and `-' signs mean
the retrograde/prograde motions. It is seen that most orbits are
located in the range between the prograde and retrograde motions;
only small number of orbits are disk orbits. For the orbit
classification, we use the method of \cite{1998MNRAS.298....1C}. In
Table~\ref{orbit_family}, we show the relative fractions of orbit
families with non-zero weights for model 23. It is seen that
irregular orbits dominate the orbit families, the fraction of these
orbits is over $90\%$ for the models if the integration time is one
Hubble time. We also find that the relative orbit fraction strongly
depends on the integrated orbit time, the fraction of irregular
orbits increases with increasing orbit integration time.

\begin{table}
\caption{$\chi^2$ for different input models, which are constrained
by the velocity and velocity dispersion in four windows $(b=-4^{\circ},
b=-6^{\circ}, b=-8^{\circ}$ and $l=0^{\circ})$.}
\label{tab:model}

\label{fit}
\begin{center}
\begin{tabular}{llllllllllllll}\hline
 model ID &$\Omega_{\rm p}$ &$M_{\rm d}$&$\theta_{\rm bar}$&${\chi^2}$\\
          &$(\rm{km\ s^{-1}\ kpc^{-1}})$ &$(M_{\rm bar}$)&$(^\circ)$& & \\

 \hline\hline
 1  &40    &2.25     &13.4 &331      \\
 2  &40    &2.25     &20   &328      \\
 3  &40    &2.25     &30   &430      \\
 4  &40    &5        &13.4 &359      \\
 5  &40    &5        &20   &393      \\
 6  &40    &5        &30   &438      \\
 7  &40    &8        &13.4 &851      \\
 8  &40    &8        &20   &902      \\
 9  &40    &8        &30   &843      \\
\hline
 10 &50    &2.25     &13.4 &363      \\
 11 &50    &2.25     &20   &340      \\
 12 &50    &2.25     &30   &272      \\
 13 &50    &5        &13.4 &297      \\
 14 &50    &5        &20   &279       \\
 15 &50    &5        &30   &284      \\
 16 &50    &8        &13.4 &606       \\
 17 &50    &8        &20   &633      \\
 18 &50    &8        &30   &561      \\
 \hline
 19 &60    &2.25     &13.4 &456      \\
 20 &60    &2.25     &20   &444      \\
 21 &60    &2.25     &30   &379        \\
 22 &60    &5        &13.4 &308     \\
 23 &60    &5        &20   &293     \\
 24 &60    &5        &30   &298      \\
 25 &60    &8        &13.4 &403     \\
 26 &60    &8        &20   &354       \\
 27 &60    &8        &30   &344      \\
 \hline
 28 &80    &2.25     &13.4 &314    \\
 29 &80    &2.25     &20   &374    \\
 30 &80    &2.25     &30   &371    \\
 31 &80    &5        &13.4 &319       \\
 32 &80    &5        &20   &398      \\
 33 &80    &5        &30   &273      \\
 34 &80    &8        &13.4 &379    \\
 35 &80    &8        &20   &352      \\
 36 &80    &8        &30   &467       \\
  \hline
 \end{tabular}
\end{center}

\end{table}
\clearpage

\clearpage

\begin{table}
\caption{Observed proper motion dispersions in some fields. The
bottom four rows are predictions from model 23.}\label{pm}
\begin{center}
\begin{tabular}{lllllllll}\hline
 Field & (l,b)&$\sigma_l$& $\sigma_b$ & Ref. \\ \hline\hline
       & $(^{\circ})$&$(\rm{mas\ yr^{-1}})$  &$(\rm{mas\ yr^{-1}})$&  \\ \hline
 Baade's Window & (1,-4)  &$3.2\pm0.1$  &$2.8\pm0.1$ & \cite{1992AJ....103..297S}\\
 Baade's Window & (1,-4)  &$3.14\pm0.11$  &$2.74\pm0.08$ & \cite{1996ApJ...470..506Z}\\
 Baade's Window & (1.13,-3.77)  &2.9  &2.5 & \cite{2002AJ....124.2054K}\\
 Baade's Window & (1,-4)  &$2.87\pm0.08$  &$2.59\pm0.08$ & \cite{2006MNRAS.370..435K}\\
 Baade's Window & (0.9,-4)  &$3.06\pm0.11$  &$2.79\pm0.13$ &\cite{2007ApJ...665L..31S}\\
 Baade's Window & (1,-4)  &$3.13\pm0.16$  &$2.50\pm0.10$ &{\cite{2010A&A...519A..77B}}\\
 Baade's Window & (1.13,-3.76)  &$3.11\pm 0.08 $  &$2.74\pm 0.13$ & Soto (2012) in preparation \\
 Plaut's Window & (0,-8)  &$3.39\pm 0.11$  &$2.91\pm 0.09$ & \cite{2007AJ....134.1432V,2009RMxAC..35..123V}  \\
 Sagittarius I & (1.25,-2.65)  &3.3  &2.7 & \cite{2002AJ....124.2054K}\\
 Sagittarius I & (1.27,-2.66)  &$3.07\pm 0.08$ &$2.73\pm 0.07$& \cite{2006MNRAS.370..435K}\\
 Sagittarius I &(1.25,-2.65)&3.067 &2.760 &\cite{2008ApJ...684.1110C}\\
 Sagittarius I & (1.26,-2.65)  &$3.56\pm 0.08 $  &$2.87\pm 0.08$ &Soto (2012) in preparation  \\
 NGC 6558 & (0.28,-6.17)  &$2.90\pm 0.11$ &$2.87\pm 0.13$ &Soto (2012) in preparation\\ \hline
 Baade's Window & (1,-4)  &4.44  &2.52 & Model 23\\
 Plaut's Window & (0,-8)  &5.28  &2.32 & Model 23\\
 Sagittarius I & (1,-3)  &4.43 &2.67 &Model 23  \\
 NGC 6558 &(0,-6) &4.46 &2.36 &Model 23\\

 \hline
 \end{tabular}
\end{center}
\end{table}

\clearpage

\clearpage

\begin{table}
\caption{Tangential velocity dispersion in some fields. The values
given in the references are derived from proper motions by assuming a
distance to the Galactic centre $R_0=8\,$kpc.}\label{pm2}
\begin{center}
\begin{tabular}{lllllllll}\hline
 Field & (l,b)&$\sigma_l$& $\sigma_b$ & Ref. \\ \hline\hline
       & $(^{\circ})$&$(\rm{km\ s^{-1}})$  &$(\rm{km\ s^{-1}}$)&  \\ \hline
 Baade's Window & (1,-4)  &$121\pm4$ &$106\pm4$ & \cite{1992AJ....103..297S}\\
 Baade's Window & (1,-4)  &$119\pm4$  &$104\pm3$ & \cite{1996ApJ...470..506Z}\\
 Baade's Window & (1.13,-3.77)  &111  &100 & \cite{2002AJ....124.2054K}\\
 Baade's Window & (1,-4)  &$109\pm3$  &$98\pm3$ & \cite{2006MNRAS.370..435K}\\
 Baade's Window & (0.9,-4)  &$116\pm4$  &$106\pm5$ &\cite{2007ApJ...665L..31S}\\
 Baade's Window & (1,-4)  &$119\pm6$  &$95\pm4$ &{\cite{2010A&A...519A..77B}}\\
 Baade's Window & (1.13,-3.76)  &$118\pm 3$  &$104\pm 5$ & \cite{2012arXiv1202.5892S} \\
 Plaut's Window & (0,-8)  &$129\pm 4$  &$110\pm4$& \cite{2007AJ....134.1432V,2009RMxAC..35..123V}  \\
 Sagittarius I & (1.25,-2.65)  &123  &108 & \cite{2002AJ....124.2054K}\\
 Sagittarius I & (1.27,-2.66)  &$117\pm 3$&$104\pm 3$& \cite{2006MNRAS.370..435K}\\
 Sagittarius I &(1.25,-2.65)&116 &105 &\cite{2008ApJ...684.1110C}\\
 Sagittarius I &(1.26,-2.65)  &$135\pm 3$  &$109\pm 3$ &\cite{2012arXiv1202.5892S}  \\
 NGC 6558 & (0.28,-6.17)  &$110\pm 4$ &$109\pm 5$ &\cite{2012arXiv1202.5892S}\\
 \hline
 Baade's Window & (1,-4)  &138  &90 & Model 23\\
 Plaut's Window & (0,-8)  &121  &71 & Model 23\\
 Sagittarius I  & (1,-3)  &145 &97 &Model 23  \\
 NGC 6558 & (0,-6)  &125 &78 &Model 23\\
 \hline
 Baade's Window & (1,-4)  &140  &106 & ZH96 \\
Sagittarius I & (1,-3)  &146&118 & ZH96  \\

 \hline
 \end{tabular}
\end{center}
\end{table}

\clearpage
\begin{table}
\caption{Relative fractions of orbit families with non-zero weights
for model 23.} \label{orbit_family}
\begin{center}
\begin{tabular}{llllllllllllll}\hline
 Orbit families & Fractions$^{a}$ & Fraction$^{b}$ &Fraction$^{c}$ &Fraction$^{d}$ \\
 \hline
irregular       &$92.9\%$ &$91.2\%$ &$81.3\%$  &$80.6\%$     \\
open z tube     &$5.8\%$  &$5.6\%$  &$13.5\%$  &$12.0\%$       \\
thin z tube     &$0.5\%$  &$2.2\%$  &$2.27\%$  &$2.6\%$        \\
others$^{e}$    &$0.8\%$  &$1.0\%$  &$2.93\%$  &$4.8\%$        \\
 \hline
\end{tabular}
\end{center}
{\footnotesize
 \noindent
 $^{a}$ The integration time of orbits is one Hubble
 time.\\
 $^{b}$ The integration time of orbits is one half Hubble
 time.\\
 $^{c}$ The integration time of orbits is one third Hubble
 time.\\
 $^{d}$ The integration time of orbits is one fourth Hubble
 time.\\
 $^{e}$ others mean closed z tube, open box, open x tube etc. \\
 }

\end{table}
\clearpage

\section{Stability of the model}

A perfect dynamical model must be self-consistent and stable. From the projected density map
and radial kinematics, we know that our model is self-consistent. However, its dynamical stability still needs to be demonstrated. For a non-spherical model, N-body simulations can help us
check the stability of the model. For this purpose, we use the versatile code
GADGET-2 \citep{2005MNRAS.364.1105S} to run our galaxy simulation.

One important step in N-body simulations is to generate the initial
conditions. In our study, we follow \cite{1996MNRAS.283..149Z} and
\cite{2009MNRAS.396..109W}. The total number of initial particles is
$2\times10^5$. They are randomly sampled from non-zero weight
orbits. In each orbit, the selected number of particles is
proportional to its orbit weight. The disk part is not in
equilibrium outside $3\ \rm{kpc}$, and $30\%$ of the disk mass
escaped during the simulation. The bar is essentially in equilibrium
with an overall drift of $2\ \rm{km\ s^{-1}}$ perhaps due to the
asymmetric escape of the disk particles.

\begin{figure}
\includegraphics[angle=0, width=80mm]{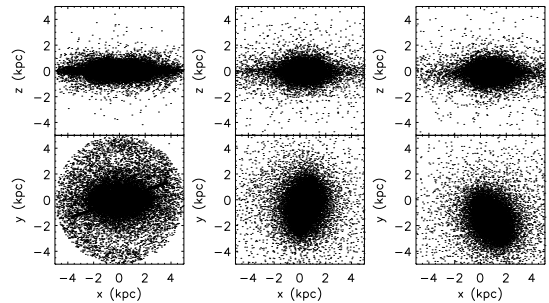}
\caption{Snapshots for the steady-state model of N-body simulations
for model 23. From left to right, the output time is 0, 0.1 and 0.5
Gyr, respectively. The solid line at the bottom left panel indicates
the line of sight to the Galactic centre.} \label{shape}
\end{figure}

The N-body simulation for our best-fit model 23 is run for 2 Gyr,
and 30 snapshots of particle positions and velocities are stored.
Figure~\ref{shape} shows the snapshots of the model at 0, 0.1 and
0.5 Gyr.

Two imperfections are noted: by construction, our disk part is not
in equilibrium outside $3\ \rm{kpc}$ where no self-consistency was
imposed in the Schwarzschild model. As a result about $30\%$ of the
disk mass evaporates due to the relatively shallow potential of the
live particles. The bulk of the disk remains, however, the
non-equilibrium of the disk makes it difficult for us to assess the
long-term stability of the bar model. Secondly the center of the bar
begins to drift slightly from the origin. The drift speed is very
low, about $2\ \rm{km\ s^{-1}}$, probably due to the recoil momentum
of asymmetric evaporation of disk particles beyond $3\ \rm{kpc}$.

Nevertheless an examination of the moment of inertia $I_{XY}$,
$I_{XX}$ and $I_{YY}$ reveals that the bar rotates about 5 times
(i.e., there are 10 peaks in $I_{XY}$, $I_{XX}$ and $I_{YY}$) in 0.5
Gyrs, which is consistent with a constant pattern rotation speed
$60\ \rm{km\ s^{-1}\ kpc^{-1}}$. The axis ratio appears stable as
well with very little evolution within 2 Gyrs. Clearly more
runs are necessary with particular attention to include a disk component.

\section{Conclusion and Discussion}

We have constructed 36 nuclear models using Schwarzschild's
method, varying the bar angle, bar pattern speed and disk mass.
Through $\chi^2$ fitting, we find the best-fit model has
$\Omega_{\rm p}=60\, {\kms\,{\rm kpc}^{-1}}$, $M_{\rm
d}=1.0\times10^{11}M_{\odot}$ and $\theta_{\rm bar}=20^{\circ}$. The
model can reproduce the three-dimensional density, surface density
and BRAVA velocity and velocity dispersion well. We tested two
different methods of smoothing and two methods of generating initial
conditions; our results are independent of these.

Compared to the model of ZH96, our model can better reproduce the
rotation curve as seen in the average radial velocity and  the
surface brightness distributions. However, our model is incomplete
in at least two aspects, the first is that the predicted proper
motions appear to be too high, and the second is the stability of
the system is far from perfect. We discuss these two issues in turn.

The proper motion dispersions in some windows predicted from the model are
higher than observed along the Galactic longitude. Possible reasons are:
\begin{enumerate}
\item The nuclear model in our paper is not perfect. Although we have
calculated 36 models with different parameters, it is possible that
we still miss the model with the right combinations. In particular,
we have used a fixed bar mass ($2.0\times10^{10}M_{\odot}$) in this
study. Moreover, the density models adopted in this paper are obtained
from fitting the surface brightness of the inner Galaxy. It has
been shown that models with different axis ratios can fit the
surface brightness of the bar \citep{1996ApJ...470..506Z}. In other
words, the three-dimensional density distribution is
not unique. Our model shows a stronger anisotropic distribution in
proper motion than in observations, thus a less
triaxial bar model may be preferred.

\item The proper motion is obtained by using the orbits inner 5kpc around the Galactic
center. If we use the orbits only inner 2.5\,kpc around the Galactic
center,  the predicted value of proper motion will be changed. For
example, in BW, $\sigma_l=3.72\rm{\,mas\ yr^{-1}}$  and
$\sigma_b=2.53\rm{\,mas\ yr^{-1}}$, then the agreement between the
model prediction and observation improves. A useful way to
compare the model prediction with those observations is to use both
the tangential velocity dispersion and proper motion dispersion. It
is seen that the predicted velocity dispersion in the Plaut's Window along
the longitude direction is close to the observed value
(See Table ~\ref{pm2}) .

\item The predictions in our model are for pixels of $1^{\circ}\times
1^{\circ}$, the observed regions are far smaller. The predicted
proper motions of model 23 in the range
$l=[-12^{\circ},12^{\circ}]$, $b=[-10^{\circ},10^{\circ}]$ are
available online \footnote{http://cosmology.bao.ac.cn/\~\,wangyg/}.
Future observations of large proper motion samples can help us to
further constrain the models.
\end{enumerate}

\begin{figure}
\includegraphics[angle=0, width=80mm]{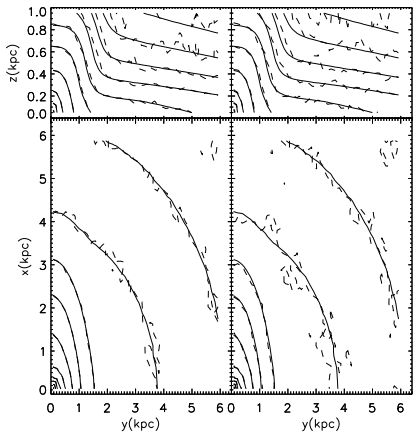}
\caption{Density contours from the input model (solid lines) and
from orbits (dashed lines) in the $x-y$ and $y-z$ planes for model
23. The left panel is for the result of the orbit weights are solved
by only using the three-dimensional density while the right one is
that the orbit weights are solved by using the three-dimensional
density, velocity and velocity dispersion.}\label{3d_6kpc}
\end{figure}

The second shortcoming of our model is the dynamical instability.
N-body simulations show that the model is only stable for 0.5 Gyr.
The initial conditions for N-body are generated from the orbit
weights. In our model, most orbits are irregular. The fraction of
irregular orbit strongly depends on the potential. Our model
includes a prolate bar, a boxy bulge and an axi-symmetric disk; the
axis ratios of these three components
are different. 
In particular, the presence of the bar implies the axi-symmetric
disk should not be present in the central part since no circular
orbits can exist. This may have limited the dynamical stability of
our system. In addition, we only consider the self-consistency of
the model within 3\,kpc. The disk is an important component in our
model which dominates the mass beyond 3\,kpc. We also check the
self-consistency of the model inside 6\,kpc, which covers the outer
Lindblad resonance region $(\sim \rm{6kpc})$. Figure~\ref{3d_6kpc}
compares the input
 and the reconstructed densities from orbits. As can be seen, the agreement
 is good within 3\,kpc, but the scatters become increasingly larger
 $(\sim 30\%)$ beyond 3\,kpc. The N-body simulation again
shows that the model is only stable within 0.5 Gyr. In the future,
it may be desirable to start with a bar model from N-body
simulations such as that from \cite{2010ApJ...720L..72S}.

At present, most density or potential models of the Galactic nucleus
are constructed using photometric data (surface brightness and star
counts) alone. However, no
density model is constructed including the kinematics, such as velocity,
velocity dispersion and proper motion. In the future, it may be useful to
construct the Galactic bar model using the density and photometry at the same time.
Future observations such as GAIA will be particularly valuable to help us construct a density model.

\section*{Acknowledgements}
We thank the referee Daniel Pfenniger for the comments and
suggestions. We acknowledge helpful discussions with Juntai Shen,
Zuhui Fan, Lia Athanassoula, Simon White, Ortwin Gerhard, Inma
Martinez-Valpuesta, and the hospitality of the Aspen Center for
Physics. We also thank other members of the BRAVA collaboration,
especially A. Kunder, who compiled the final public release dataset,
and  M. Soto, who provided us the unpublished proper motion data.

This work was started during Y.G.W.'s visit in 2008 to the Jodrell
Bank Centre for Astrophysics supported by their visitor's grant.
Y.G.W. acknowledges the support by the National Science Foundation
of China (Grant No. Y011061001 and No. Y122071001). and S.M. thanks
the Chinese Academy of Sciences for financial support. R.M.R.
acknowledges support from NSF grant AST0709749, and GO-11655.01.

Most of the computing was performed on the supercomputer ``laohu" at
the High Performance Computing Center at National Astronomical
Observatories of China, funded by Ministry of Finance under the
grant ZDYZ2008-2.

\bibliographystyle{mn2e}
\bibliography{bar}

\appendix

\label{lastpage}

\clearpage
\end{document}